# Integration of a $^6$LiInSe$_2$ Thermal Neutron Detector into a CubeSat Instrument

**Joanna C. Egner,**[a,b] **Michael Groza,**[b] **Arnold Burger,**[a,b] **Keivan G. Stassun,**[a,b] **Vladimir Buliga,**[b] **Liviu Matei,**[b] **Julia G. Bodnarik,**[c] **Ashley C. Stowe,**[a,d] **Thomas H. Prettyman**[e]

[a] Department of Physics and Astronomy, Vanderbilt University, Nashville, TN 37240
[b] Department of Physics, Material Science and Application Group, Fisk University, Nashville, TN 37208
[c] Lunar and Planetary Laboratory, The University of Arizona, Tucson, AZ 85721
[d] Y-12 National Security Complex, Oak Ridge, TN 37830
[e] Planetary Science Institute, Tucson, AZ 85719

**Abstract.** We present a preliminary design for a novel neutron detection system that is compact, lightweight, and low power consuming, utilizing the CubeSat platform making it suitable for space-based applications. This is made possible using the scintillating crystal lithium indium diselenide ($^6$LiInSe$_2$), the first crystal to include $^6$Li in the crystalline structure, and a silicon avalanche photodiode (Si-APD). The schematics of this instrument are presented as well as the response of the instrument to initial testing under alpha radiation. A principal aim of this work is to demonstrate the feasibility of such a neutron detection system within a CubeSat platform. The entire end-to-end system presented here is 10 cm x 10 cm x 15 cm, weighs 670 grams and requires 5 Volts direct current at 3 Watts.

**Keywords**: neutron detection instrument, CubeSat, scintillator, Si-APD.

## 1 Introduction

There is a growing interest in the characterization of near-Earth objects such as the moon, asteroids, and planets due to an expansion of investigation of the extra-terrestrial for both scientific and economic purposes. Characterization includes the investigation of these objects' bulk composition, in effort to better understand the chemistry of the sub-surface, which in turn could give constraints on solar system formation theories and furthermore provides potentially valuable commercial information such as asteroid mining. Previous nuclear measurements have primarily given information regarding the composition of the top-most surface layer, on the order of about 10 centimeters [1]. While the characterization of the surface of objects is valuable, often the surface and bulk compositions are very different because of exposure and weathering effects that predominantly affect the surface such as ionization from the sun, the evaporation of liquids due to low surface gravity or lack of atmosphere, and erosion from cosmic wind [2].

Historically, the classification of asteroids has been carried out by optical albedo observations, measuring the scintillation spectrum of light that they give off [3]. However, because the bulk and surface composition can differ greatly, albedo measurements are not always the most accurate way of determining composition of the bulk of the asteroid. With the development of a better system of elemental detection, classification could be done much more accurately. Not only would this provide important information that could help answer questions pertaining to the origin and distribution of asteroids, it would also have major implications in the developing sector of asteroid mining. Making the classification and identification of the best candidates for mining easier and more accurate. A lithium indium diselenide ($^6LiInSe_2$) based CubeSat instrument will have the capability to probe to depths of 50 centimeters below the surface of solar system bodies, while weighing and consuming a fraction of the weight and power of the traditional instruments described in the literature [1, 4, 5, 6, 7]. The use of neutron detectors will be a major component in studies of the chemical composition of asteroids and planets.

Previously all neutron-detecting instruments have been large and require high voltage. Therefore it would have been impossible to package one of these detectors keeping in the boundaries of the CubeSat platform. Examples of neutron detectors included on previous missions include the Helium-3 tube detector included in the 1998 Lunar Prospector, the boron-loaded plastic scintillator included in the 2001 Mars Odyssey mission and the 2007 gamma ray and neutron detector (GRaND) Dawn mission to explore Ceres and Vesta which utilized boron-loaded plastic and lithium-loaded glass scintillators, as well as a semiconducting cadmium zinc tellurium (CdZnTe) crystal array [5, 8, 6]. These instruments are both heavy and have high power requirements. 30 kg, 32 W for Mars Odyssey and 10 kg, 15 W for GRaND [5, 8]. While $^3$He gas tubes are highly efficient neutron detectors due to a high cross-section for neutron detection, they are heavy, require high voltage, and are bulky. Similarly, most scintillator-based detectors require pairing with photomultiplier tube (PMT) systems, which are sensitive to magnetic fields and add weight to payload and increase power consumption. Therefore there is great interest in developing a solid-state neutron detector that is more compact, requires less power, and does not have sensitivity to magnetic fields to implement into a cost-effective space instrument.

$^6$Li has a large cross-section, 940 barns, for thermal neutron capture following the reaction $^6$Li + $^1$n $\longrightarrow$ $^7$Li* $\longrightarrow$ $^4$He(2.056MeV) + $^3$H(2.729MeV), the Q-value, or the energy released, depends on the energy liberated following the neutron capture and in this case is 4.785 MeV [9]. Traditionally, $^6$Li is used as an absorber on semiconductors to increase efficiency but $^6$LiInSe$_2$ the first semiconductor to have $^6$Li in the structure of the crystal. It can be operated as either a semiconductor or a scintillator, at the current state of development and purification $^6$LiInSe$_2$ is more efficient to use as scintillator [10, 11, 12]. Since $^6$LiInSe$_2$ can be used in either semiconductor or scintillator mode, one of the initial decisions we had to make was the preferred mode of operation for the CubeSat prototype. We have found that, at the present level of material development, the signal-to-noise ratio is improved by 42% in the scintillator mode of operation.

The development of a lithium indium diselenide $^6$LiInSe$_2$ [10] thermal neutron sensor is especially attractive because of its ability to be spectrally matched (510 nm emission) with newly developed silicon based photodetectors such as silicon avalanche photodiodes (Si-APD), and silicon photomultipliers (SiPM), that can be used in place of traditional photomultiplier tubes. These are much less bulky and heavy, and have the potential to be packaged into a compact, lightweight, low power detector system. Comparing models from manufacturer Hamamatsu Photonics, their SiPM weights 1g and their PMT weights 100g, the SiPM is also 10% smaller.

In this paper the prototype of a neutron detection CubeSat system and the preliminary performance tests that demonstrate the ability to deploy a neutron detection instrument in a compact footprint are discussed. The prototype combines the innovative technology of $^6$LiInSe$_2$ and a Si-APD. This pairing eliminates the need for a conventional photomultiplier tube making it possible to package this instrument under the restrictions of the CubeSat program. Section 2 describes the design of the instrument and briefly discusses each component as well as the reasoning behind selection. Section 3 presents preliminary results and the basics of the instrument functions.

## 2  Instrument Design

### 2.1  CubeSat Design Specifications

NASA's CubeSat Launch Initiative is a platform that allows researchers to compete for flight opportunities to conduct low cost space science experiments. Instruments can be from one to six units (U), with some sub-increments of 0.5 units available, with specific weight, size, and power requirements that can be found on www.cubesat.org. The criteria for a 1.5U include the size of 10 cm x 10 cm x 15 cm, a weight limit of 2.00 kg (4.4 lbs), and requirement to run off the power supplied by onboard solar panels or batteries [13]. The 0.5U IRIS Deep Space Transponder CubeSat, built by JPL, has a power consumption of 26 W during full power transmission, and the 6U JPL MarCO, has a power consumption of 35 W  [14, 15].

2.2     CubeSat Design

In the development of our instrument we have utilized a 1.5 unit CubeSat chassis. A block diagram of the instrument is shown in *Figure 1*. In this figure, the active area of the detector, the encapsulated $^6$LiInSe$_2$ crystal, is oriented to be at the top. A thermal neutron interacts with the $^6$Li nucleus, and the resulting nuclear reaction generates an energetic alpha particle that in an ionization cascade generates excited states in the scintillator crystal that ultimately recombine to their ground state by emitting optical photons [9]. The scintillating photons are then detected by the Si-APD and converted into an electronic signal, or pulse, that is further processed to generate a histogram of all events that constitute the spectrum of incoming radiation. In general, neutron detection takes place in a mixed field of other ionizing particles such as gamma and galactic cosmic rays and the characteristics of the pulse can be used to discriminate between types of particles interacting with the detector.

*Figure 1* shows the schematic of our instrument. The instrument has three stages, comprising of a printed circuit board (labeled Boards 1, 2, and 3 in all figures) and the associated components. The first board contains the 9 mm x 9 mm x 2 mm $^6$LiInSe$_2$ encapsulated crystal and the Si-APD (components 1 and 2 in all figures), which have been packaged together to prevent the crystal from shifting off the active area of the Si-APD that is slightly larger than the crystal at 10x10 mm$^2$. Packaging the crystal and Si-APD together also decreases the amount of ambient light that can reach the Si-APD. Board 2 contains two DC-DC regulated power supplies (components 5 and 6) that convert the 5 Volt input to the instrument into 12V and 440V respectively. The 440 V power supply powers the Si-APD, while the 12 V supply runs the remaining components, the two amplifiers and the Kromek multichannel analyzer (MCA).

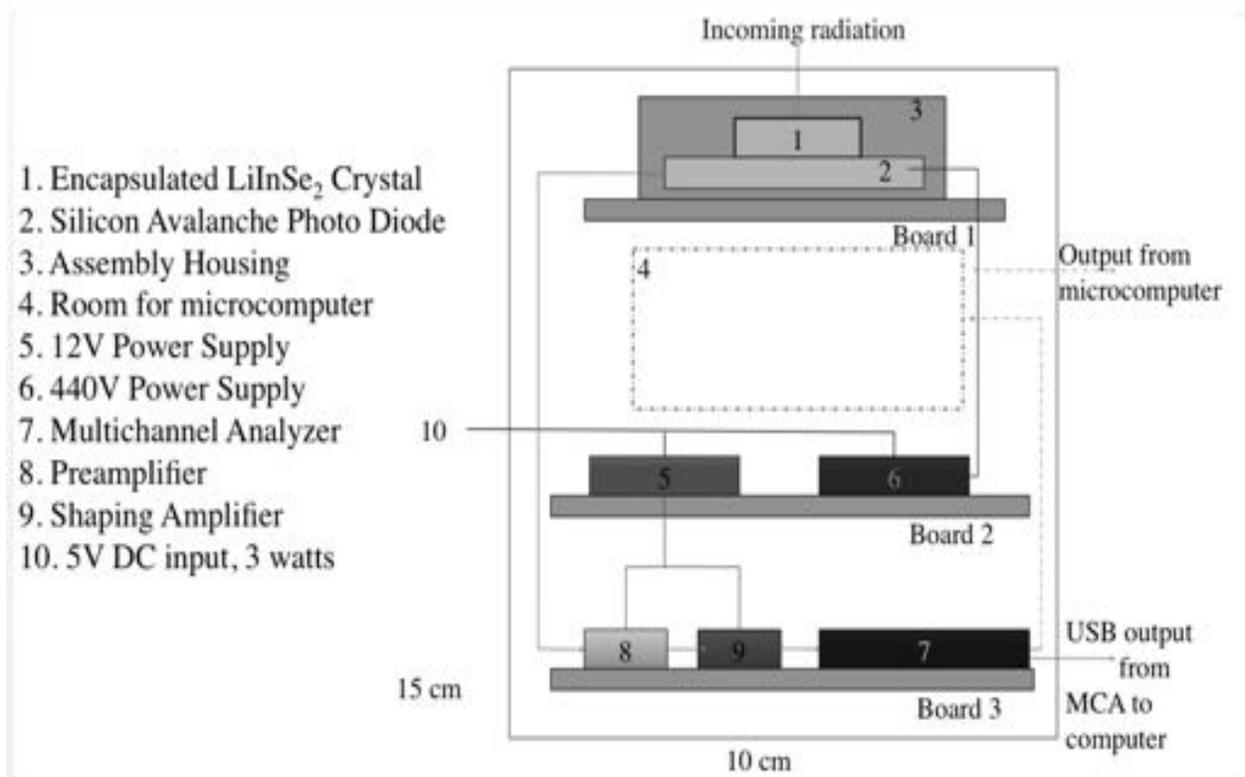

**Figure 1:** Neutron CubeSat prototype schematic

The third board contains a preamplifier and shaper (component 8) and a voltage amplifier (component 9) and the MCA (component 7), which currently, for lab tests, outputs via USB cable to a computer. There is room as well as weight and power available for a microcomputer and wireless transmitter to be included in a later iteration of the instrument, though a RaspberryPi is being used in initial tests. The current instrument, including the RaspberryPi microcomputer weighs 670 grams in its entirety, and requires 5V 3W.

The $^6$LiInSe$_2$ crystal being used in the detector was grown using the vertical Bridgman method with details of growth being given by Tupytsin *et al.* and using a technology that was developed through a partnership between Fisk University and Y-12 National Security Complex [10]. $^6$Li is a high-density material suitable for producing lightweight and compact neutron detectors due to its high capture cross section for thermal neutrons. $^6$Li and other neutron sensitive materials are traditionally used as coatings on neutron detecting systems to increase the efficiency to a total of a few percent, in comparison, the density of $^6$Li in $^6$LiInSe$_2$ is great enough to yield up to 95% detection efficiency of neutrons in a 3.4 mm thick wafer [11].

The $^6$LiInSe$_2$ crystal utilized in the detector is 9mm x 9mm x 2mm. After characterization and testing, discussed more in section 3, it was encapsulated, covered on five sides with reflective material with a quartz window on the sixth side, shown in the topmost panel of *Figure 2* labeled 1, this side was then placed onto the Si-APD.

The photo detector, in this case the Si-APD (component 2) turns detected events into electrical pulses. Compared to Si-APDs and silicon based photomultipliers (SiPMs), traditional vacuum tube photomultipliers (PMTs) are heavy, large, have high power requirements, and are sensitive to the magnetic environment of space. For these reasons the use of a silicon based photo detector is better suited for space applications. In this instrument a

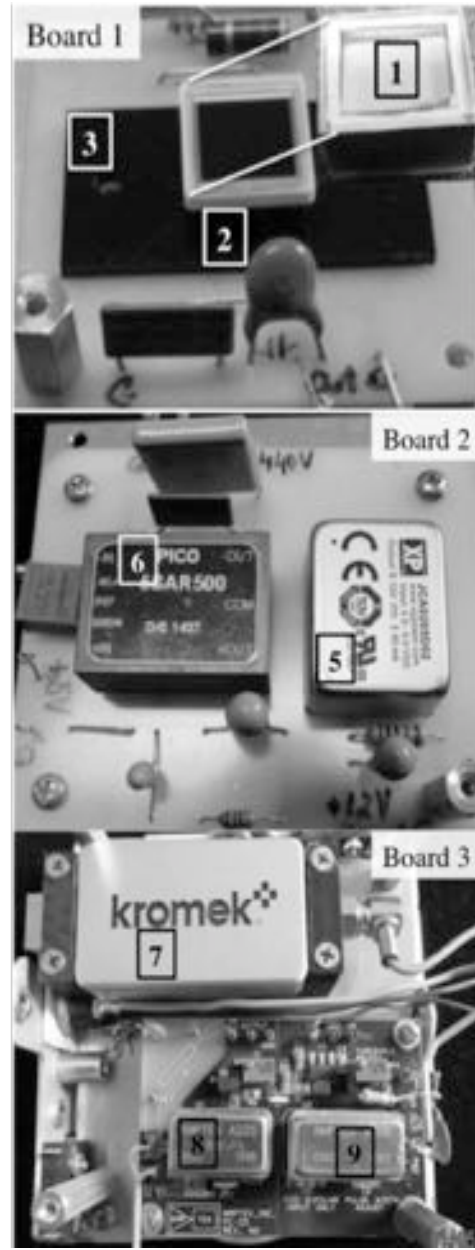

**Figure 2**: Photo of associated electronics. Labeling of components is the same as the one used in Fig.1

Hamamatsu S8664 1010 Si-APD (component 2) was utilized. APDs utilize the photoelectric effect to turn the scintillated light from the crystal into an electronic signal [9].

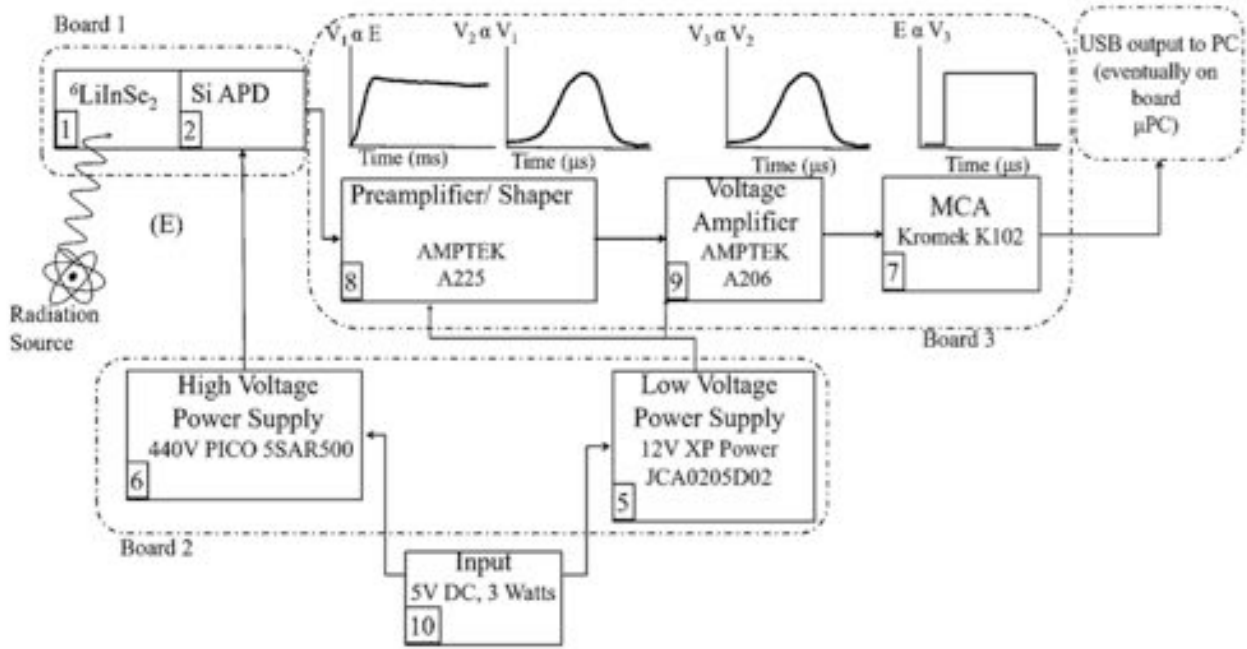

**Figure 3:** Block diagram of the detection system including LiInSe$_2$ crystal (1), Si-APD (2), associated electronics for signal processing (7,8,9) and power supplies (5 and 6). The main information extracted from each signal is the height of the pulse, which is linearly proportional to the energy of the ionizing particle detected. Labeling of components is the same used in previous figures.

Next the preamplifier and shaper, here an Amptek A225 (component 8), receives the signal from the Si-APD, and preserves the energy of the signal in a pulse that initially has the following characteristics: a sharp rise and a long tail with total signal on the order of micro seconds (shown in *Figure 3*). Next, because the Amptek A225 acts as a shaper as well, it then takes the signal and cuts down the pulse length to the order of microseconds, preserving the energy as the height of the pulse and changing the shape of the signal to be closer to a Gaussian shape. Next the voltage amplifier, an Amptek 206 (component 9), amplifies the voltage to the pulse. Then the signals outputted by the voltage amplifier then go to the multi channel analyzer (MCA), a Kromek K102 (component 7). The MCA measures the Gaussian peaks from the voltage amplifier by pulse height (with the energy value still represented by height – shown in *Figure 3*) and sorts pulses by their energy creating an energy spectrum of the interacting events.

## 3 Results and Discussion

### 3.1 $^6$LiInSe$_2$ Crystal Scintillator

The response to alpha particles is a good indication of the performance of the system before neutron testing. The $^6$LiInSe$_2$ crystal was fabricated into a scintillator and its response to alpha radiation from a $^{241}$Am source with an activity of 0.9 μCi was tested at room temperature before encapsulation and integration into the instrument. The results are shown in *Figure 4*. $^{241}$Am decays via the emission of an alpha particle with energy 5.486 MeV to $^{237}$Np (Neptunium) [16]. The energy resolution of 36% was measured as the full width of the distribution at half the maximum of the peak (FWHM). This demonstrates a relatively high signal-to-noise (or signal-to-background) ratio of our system, approximately we will be able to distinguish between two energies separated by 36% or more [9]. The low energy tail that can be seen in *Figure 4* is due to general background, that includes the low energy, 60 KeV, gammas that are generated by this source.

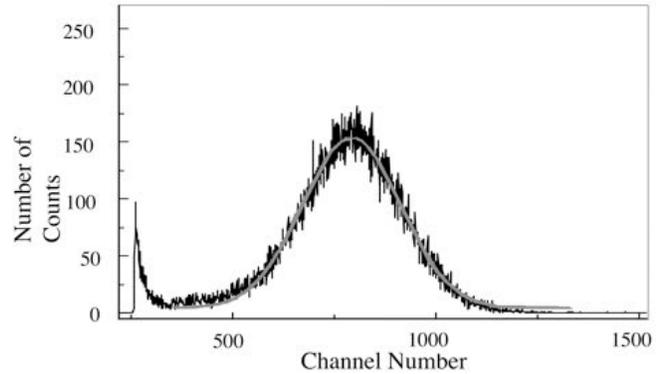

**Figure 4:** $^6$LiInSe$_2$ crystal response to alpha radiation from a $^{241}$Am source. The spectrum was collected over 100 seconds.

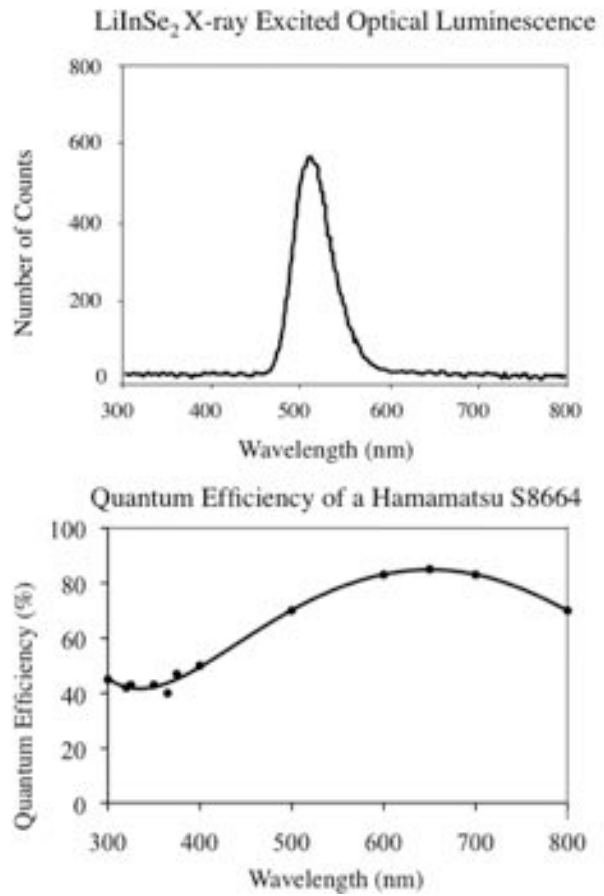

**Figure 5:** Comparison of Quantum Efficiency of a Hamamatsu S8664 1010 Si-APD and the output spectra of a $^6$LiInSe$_2$ crystal

3.2 Silicon Avalanche Photodiode (Si-APD)

Available scintillators emit in the blue or ultraviolet parts of the spectrum and therefore do not pair as well with silicon-based photo-detectors as $^6$LiInSe$_2$ does [12, 17, 18]. This is shown in *Figure 5*, where the quantum efficiency of the Hamamatsu S8664, given by the manufacturer, and the emission spectrum of a $^6$LiInSe$_2$ crystal, collected using X-ray excited optical luminescence, are shown. It can be seen that the $^6$LiInSe$_2$ crystal has its peak emission at about 510 nm, and that the Si-APD has a high efficiency in this range. The blue or UV part of the spectrum, where most other scintillators emit, corresponds to the 400 nm range and lower. From 500 to 400 nm, there is a drop of efficiency from 80 to 40% of the Si-APD. High efficiency in the photo-detector is important because $^6$LiInSe$_2$ does not have a high light yield, 4,400 photons/MeV compared to inorganic and plastic scintillators which have light yields in the 10,000s (for example CsI(Tl) has a light yield of 65,000 photons/MeV) [12, 9].

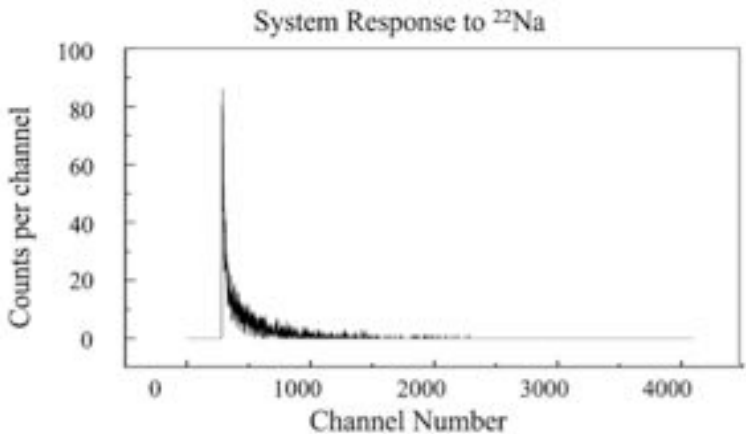

**Figure 6:** Response of packaged instrument to gamma radiation from a $^{22}$Na source. Of the 5095 total counts collected, 70% are from direct interaction with the Si-APD.

3.3 Post-assembly System Test

Once the instrument was built, it was tested under gamma radiation. After encapsulation the system was tested by measuring the response to the 1.2 MeV gammas emitted by [22]Na, the spectrum collected is shown in *Figure 6.* The height of the [22]Na gamma signal is significantly lower than the response measured to the alpha particles from a [241]Am source.

## 4 Conclusions and Future Work

In this study we have designed and demonstrated a prototype of a neutron detection system packaged into a 1.5U CubeSat. Weighing 670 grams it can operate on 5 V and a total power of 3 W, which is considered low, could be supplied by solar panels, and reduces the complexity required for a heat dissipation system. The instruments we are comparing to are 10-30kg, and require 15-32W (GRaND and Mars Odyssey)[5,8].

Future steps will include environmental test measurements of the $^6$LiInSe$_2$ and Si-APD assembly with a neutron source at the Lunar and Planetary Lab at the University of Arizona and validation of the results using the Monte Carlo N-Particle Transport Code (MCNP) software package. We are finalizing the experimental setup for epithermal and thermal neutron exposure in a mixed neutron and gamma-ray environment using a Cf-252 source and we will develop an MCPN model that matches the final geometry. This study will be reported in a near future publication. During these tests we will utilize the ability of the microcomputer to broadcast data using an antenna to avoid the use of lengthy cables and allow immediate access to collected data while the instrument is under shielding. Testing and characterization of the ability of the instrument to discriminate against gamma-ray events by pulse shape will also be a focus. Our long-term goals for this instrument are to make it self-contained by including a microcomputer, and to make it suitable for the environment of space by replacing **any** components that are not suited for space with radiation hard electronics [17]. This proof of concept instrument was developed and built using only "off the shelf" components.

*Acknowledgements*

For helpful feedback and discussion throughout the duration of this project, the authors would like to thank the members of Dr. Burger's Materials Science and Applications Group and Dr. Stassun's Astromaterials Group. We also acknowledge the financial support provided by the National Science Foundation through grant #AST-0849736 (PAARE program)